\documentclass[twocolumn,showpacs,prl]{revtex4}
\usepackage{bm}
\usepackage{epsfig}
\newcommand{\nix}[1]{}

\begin{document}

\title{Can an electric current orient spins in quantum wells?}
\author{  S.D.~Ganichev$^{1,2}$, S.N.~Danilov$^{1}$,
Petra~Schneider$^1$, V.V.~Bel'kov$^{2}$, L.E.~Golub$^2$,
W.~Wegscheider$^1$, D.~Weiss$^1$, W.~Prettl$^1$}
\affiliation{$^1$Fakult\"{a}t Physik, University of Regensburg,
93040, Regensburg, Germany}
\affiliation{$^2$A.F.~Ioffe Physico-Technical Institute, Russian
Academy of Sciences, 194021 St.~Petersburg, Russia}

\date{\today}

\begin{abstract}
A longstanding theoretical prediction is the orientation of spins
by an electrical current flowing through low-dimensional carrier
systems of sufficiently low crystallographic symmetry. Here we
show by means of terahertz transmission experiments through
two-dimensional hole systems a growing spin orientation with an
increasing current at room temperature.
\end{abstract}

 \maketitle

The manipulation of the spin degree of freedom in electrically
conducting systems by electric and/or magnetic fields is at the
heart of semiconductor spintronics~\cite{spintronicbook02}. Spin
control in low-dimensional systems is particularly important for
combining magnetic properties with the versatile electronic
characteristics of semiconductor heterojunctions. The feasibility
to orient the spin of charge carriers in GaAs based quantum wells
by driving an electric current through the device was
theoretically predicted more than two decades
ago~\cite{aronov89,Edelstein99p233,Vasko1982}. A direct
experimental proof of this effect is  missing so far.

In this Letter we demonstrate by means of terahertz transmission
experiments
 that an electric current which flows through a
low-dimensional electron or hole system leads to a stationary spin
pola\-rization of free charge carriers. Microscopically the effect
is a consequence of spin-orbit coupling which lifts the
spin-degeneracy in ${\bm k}$-space of  charge carriers together
with spin dependent relaxation.

\begin{figure}
\centerline{\epsfxsize 70mm \epsfbox{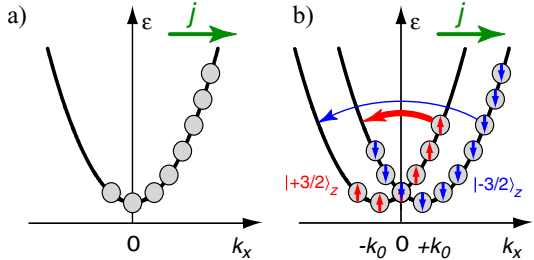}}
\caption{Comparison of current flow in (a) spin-degenerate and (b)
spin-split subbands. (a) Electron distribution at a stationary
current flow due to acceleration in an electric field and momentum
relaxation.
(b) Spin polarization due to spin-flip scattering. Here only
$\sigma_z k_x$ term  are taken into account in the Hamiltonian
which splits a valence subband into two parabolas with spin-up
(+3/2) and spin-down (- 3/2) in  $z$-direction. Biasing along
$x$-direction causes an asymmetric in ${\bm k}$-space occupation
of both parabolas. } \label{fig01}
\end{figure}

In the simplest case the electron's (or hole's) kinetic energy in
a quantum well oriented perpendicularly to  the $z$-direction
depends quadratically on the in-plane wave vector components $k_x$
and $k_y$. In equilibrium, the spin degenerated  $k_x$ and $k_y$
states are symmetrically occupied up to the Fermi energy $E_F$. If
an external electric field is applied, the charge carriers drift
in the direction of the resulting force. The carriers are
accelerated by the electric field and gain kinetic energy until
they are scattered. A stationary state forms where the energy gain
and the relaxation are balanced resulting in a non-symmetric
distribution of carriers in  ${\bm k}$-space. This situation is
sketched in Fig.~\ref{fig01}a for holes,  a situation relevant for
the experiments  presented here. The holes acquire the average
quasi-momentum
\begin{equation}\label{momentum}
    \langle {\bm k} \rangle = {e \tau_p \over \hbar}{\bm E} = {m^*  \over e \hbar p} {\bm j},
\end{equation}
where ${\bm E}$ is the electric field strength, $\tau_p$ the
momentum relaxation time, ${\bm j}$ the electric current density,
$m^*$ the effective mass, $p$ the hole  concentration and $e$ the
elementary charge. As long as spin-up and spin-down states are
degenerated in ${\bm k}$-space the energy bands remain equally
populated and  a current is not accompanied by spin orientation.
In QWs made of zinc-blende structure material like GaAs, however,
the spin degeneracy is lifted due to lack of inversion symmetry
and low-dimensional quantization
~\cite{Bychkov84p78,Dyakonov86p110} and the resulting
dispersion reads
\begin{equation}
\varepsilon = \frac{\hbar^2 \bm{k}^2}{2m^*} \pm \beta |{\bm k}|
\end{equation}
with the spin-orbit coupling strength $\beta$. The corresponding
dispersion is sketched in Fig.~\ref{fig01}b. The parabolic energy
band splits into two subbands of opposite spin direction shifted
in ${\bm k}$-space symmetrically around ${\bm k} = 0$ with minima
at $\pm k_0$.
 In the presence of an in-plane electric field the
${\bm k}$-space distribution of carriers gets shifted yielding an
electric current. Due to the band splitting carrier relaxation
becomes spin dependent. Relaxation processes including spin flips
are different for the two subbands because the quasi-momentum
transfer from initial to final states is
different~\cite{Averkiev02pR271}. In Fig.~\ref{fig01}b the $\bm
k$-dependent spin-flip scattering processes are indicated by
arrows of different lengths and thickness. As a consequence
different numbers of spin-up and spin-down carriers contribute to
the current causing a stationary spin orientation.

For the coupling constant $\beta$ and the mechanism depicted in
Fig.~\ref{fig01}b we consider solely spin orbit coupling due to a
Hamiltonian of the form  $H_{SO} = \beta \sigma_z k_x$  with the
Pauli matrix $\sigma_z$. This corresponds to a subband splitting
for eigenstates with spins pointing in $z$-direction, normal to
the quantum well plane and detectable in experiment. In our QWs of
C$_s$  symmetry the $x$-direction lies along [1${\bar 1}$0] in the
QW plane. For the moment we assume that the origin of the current
induced spin orientation is, as sketched in Fig.~\ref{fig01}b,
exclusively due to scattering and hence dominated  by the
Elliot-Yafet spin relaxation time~\cite{Averkiev02pR271}.

In order to observe current induced spin polarization we study
transmission of terahertz radiation through devices containing
multiple hole QWs. A spin polarization in $z$--direction affects,
in principle, incoming linearly polarized radiation by two
mechanisms: i) dichroic absorption and ii) Faraday rotation. The
first mechanism is based on different absorption coefficients for
left and right circularly polarized light while the Faraday
rotation is due to different coefficients of refraction for left
and right circularly polarized radiation. In experiment we used
direct inter-subband transitions between the lowest heavy-hole and
light-hole subbands of the  valence band excited by linearly
polarized terahertz radiation of a far-infrared laser. The
linearly polarized light  can be thought of being composed of two
circularly polarized components of opposite helicity.

The resulting different absorption coefficients for left and right
circularly polarized light changes the light's state of
polarization. In particular, linearly polarized radiation gets
elliptically polarized. The Faraday rotation, in contrast, becomes
important for weak absorption and is proportional to the
difference of the indices of refraction for left and right
circularly polarized radiation. In this case only the phases of
left and right circularly polarized light are shifted resulting in
a rotation of the polarization axis of the incoming linearly
polarized light. Without spin orientation in the lower subband,
the absorption strength as well as the index of refraction for
right- and left-handed polarized light are equal and transmitted
light does not change its state of polarization. However, Faraday
effect and dichroic absorption proof current induced spin
polarization.

As material we have chosen  $p$-type GaAs QWs of
 low symmetry having only - in addition to identity - one
plane of mirror reflection (i.e. C$_s$ point group according to
Sch{\"o}nflies's notation). This  was achieved by growing
modulation doped QWs on (113)A- or miscut (00l)-oriented GaAs
substrates (tilt angle: 5$^\circ$ towards the [110] direction) by
molecular-beam-epitaxy (MBE) or
metal-organic-chemical-vapor-deposition (MOCVD), respectively. Two
types of samples were prepared. Sample A: (113)A with  QW of width
$L_W$ = 10 nm,  and a free carrier density of $p\approx  2 \cdot
10^{11}$ cm$^{-2}$ and sample B: miscut (001) with $L_W$ = 20 nm
and $p \approx  2 \cdot 10^{11}$ cm$^{-2}$.
 To cope with the small
absorption signal and/or rotation angle of an individual quantum
well we fabricated multiple QW structures. Sample A  contained $N=
100$
 and sample B $N=400$ QWs, respectively. The sample
edges were oriented along  [1${\bar 1}$0] in the QW plane
($x$-direction) and perpendicular to this direction
($y$-direction). Two pairs of ohmic contacts were centered along
opposite sample edges of 5~mm width. In addition samples
containing 100 QWs and having very thin barriers were taken as
quasi-bulk reference samples.

\begin{figure*}
\centerline{\epsfxsize 120mm \epsfbox{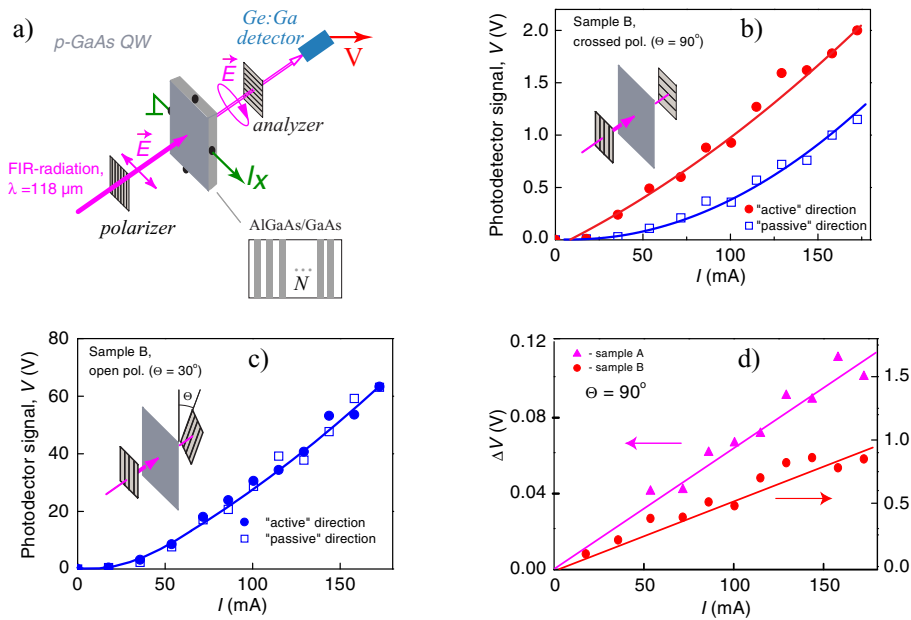}}
\caption{ (a) Experimental set-up. The sample is placed between
crossed polarizer and analyzer blocking optical transmission at
zero current through the sample. Injecting a modulated current in
the sample yields a signal at the detector which is recorded by
box-car technique. (b) and (c) Detector signal as a function of
the current in the active (full symbols) and the passive (open
symbols) directions  for sample B and two states of the analyzer:
(b) crossed polarizers ($\Theta = 90^\circ$) and  (c) partially
open polarizers ($\Theta = 30^\circ$). (d) Differences $\Delta V$
between  the signal at current in active and passive direction as
a function of current for two samples, A (triangles, left axis)
and B (circles, right axis) for crossed polarizers.  }
\label{fig03}
\end{figure*}

A spin polarization is not expected for all current directions.
For materials of low symmetry, used here, only an electric current
along $x \parallel [1{\bar 1}0]$-direction is expected to align
spins; in contrast, current flowing in $y$-direction does not
yield a spin orientation. By symmetry arguments it is
straightforward to show that a current $j_x$, in the plane of the
QW yields an average spin polarization $S_{z}$ normal to the QW
according to
\begin{equation}
S_{z} = R_{zx} j_{x} \label{eq3}
\end{equation}
where $\bm R$ is a second rank pseudo--tensor~\cite{Aronov91p973}.
However, for a current flowing along $y$-direction, $S_{z}$=0
holds since, due to symmetry, $R_{zy}$=0. Thus we expect to
observe a spin polarization for current flow in one but not in the
other (perpendicular) direction. Below we denote these directions
as active and passive, respectively.

The transmission measurements using linearly polarized 118~$\mu$m
radiation of an optically pumped $cw$ far-infrared laser were
carried out at room temperature.  The electric current (0 to
180~mA) was applied as 10~$\mu$s long pulses with a repetition
rate of 20~kHz. The schematic experimental set up is shown in
Fig.~\ref{fig03}a: the sample was placed between two metallic grid
polarizers and the $cw$ terahertz radiation was passed through
this optical arrangement (see Fig.~\ref{fig03}a). The transmitted
radiation was detected using a highly sensitive Ge:Ga extrinsic
photodetector operated at 4.2~K.

In order to detect a current dependent change of the polarization
of the transmitted light we used a crossed polarizer set-up. The
crossed polarizers are expected to let pass only light whose state
of polarization was changed by the current through the sample. The
experimental result of the transmission, which is proportional to
the photodetector signal, is shown in Fig.~\ref{fig03}b as
function of the current strength, $I$, for both  passive and
active directions. Although the signal in the active direction is
by a factor 2-3  higher than for the passive one, the transmission
signal increases in both cases with $I$. As will be pointed out
below the observed transmission for the 'passive' case is a
polarization independent background signal while the difference of
transmission between the 'active' and the 'passive' traces is the
sought-after   polarization dependent transmission signal proofing
current induced spin polarization in QWs.

To ensure that the signal for current flow in the active direction
is indeed due to spin orientation we carried out two additional
experiments. First we tested the quality of our polarizers.  As
result we obtained even for crossed polarizers ($\Theta $ =
90$^{^{\circ}})$ that a small fraction $\alpha_{90^\circ}=5.4
\cdot 10^{-3}$ of the radiation is still transmitted though we
used far-infrared polarizers of highest available quality. The
signal increasing with increasing current along the passive
direction is ascribed to carrier heating by the current. By this
process the subband hole distribution is changed and the
transmission increases with increasing
current~\cite{Vorobjev96p977}. The heating induced enhanced
transmission with increasing current together with the finite
transmission through crossed polarizers explains the nonlinear
increase of the transmission signal for current in the passive
direction (see Fig.~\ref{fig03}b). The signal for the active
direction, also displayed in Fig.~\ref{fig03}b, is markedly higher
for crossed polarizers. In a second experiment the analyzer is
rotated away from 90$^\circ$ and the signals for passive and
active direction become equal, documented in Fig.~\ref{fig03}c.
This is due to the fact that the heating induced signal increases
drastically for open polarizers whereas the signal induced by the
polarization change varies only slightly. The heating induced
signal dominates for open polarizers, whereas the polarization and
the heating induced signals, are comparable for crossed
polarizers. The purely spin polarization induced signal can be
consequently extracted from the transmission difference of  active
and  passive directions for crossed polarizers.

The difference signals for sample A and B are shown in
Fig.~\ref{fig03}d. The difference signal, reflecting the build up
of spin polarization with increasing current, increases  almost
linearly. Control experiments on the quasi-bulk sample give -- in
accordance with
 theory which forbids current induced spin
orientation for T$_d$ point group symmetry -- the same signal for
passive and active directions.

While the experiment displays clear spin polarization due to the
driving current, it is not straightforward to determine the value
of spin polarization. Due to lack of compensators  for the far
infrared regime it is difficult to judge whether the transmitted
signal is linearly (Faraday effect) or elliptically polarized
(dichroic absorption). In case of dominating dichroic absorption
the average spin polarization of a quantum well is given
by~\cite{tobepublished}
\begin{equation}
\langle S \rangle = \Delta p/p = 8 \sqrt{\alpha_{90^\circ} \Delta
V / V^{(p)}}/K_0 \,\,\,\, . \label{dichro}
\end{equation}
Here, $\Delta p$   is the difference of spin-up and spin-down hole
densities, $\Delta V$ is the spin induced photosignal plotted in
Fig.~\ref{fig03}d, and $V^{(p)}$ is the photodetector signal
obtained for a current in the passive current direction, plotted
for sample B in Fig.~\ref{fig03}b. The absorption  $K_0$, which
determines the ratio of incoming ($I_0$) and transmitted ($I_T$)
intensity through the multi-QW structure, $I_0/ I_T = \exp(-K_0)$,
is obtained from an independent transmission experiment, carried
out on unbiased devices. For sample A we obtained $K_0=2.7$, for
sample B,  $K_0= 3.4$. The would result in  spin polarization of
0.12 for sample A and 0.15 for sample B at current densities
3~mA/cm and 0.75~mA/cm per QW, respectively. If the increased
signal, however, is due to Faraday rotation a different analysis
has to be applied. The angle of Faraday rotation can be determined
by rotating the analyzer for current along the passive direction
until the signal becomes equal to the signal obtained for the
current in active direction for crossed polarizers. We obtain a
rotation angle $\varphi $ of 0.4~mrad per quantum well for sample
A and 0.15~mrad for sample B. In case of dominating Faraday
effect, however, no straightforward way to extract the value of
the spin polarization from the Faraday rotation angle is at hand.

According to the theory of Aronov et al.~\cite{aronov89} current
should yield a spin polarization on the order of $\langle
$S$\rangle \quad \approx \quad \beta \cdot \langle {k} \rangle /
k_{\rm B}T$. Using Eq.~\ref{momentum} we estimate this value  as
\begin{equation}
\label{Srel} \langle S\rangle =  \frac{Q\beta}{k_{\rm B}
T}\cdot\frac{m^*}{e\hbar  p} j,
\end{equation}
where $Q \simeq 1$  is a constant  determined by  momentum
scattering and the spin relaxation mechanism~\cite{Aronov91p973}.
For a situation where Fermi statistic applies the factor $k_{\rm
B}T$ needs to be replaced by  $2 E_{\rm F} / 3$. Calculating
$\langle S\rangle$ from Eq.\,\ref{Srel} with the experimental
parameters $p=2 \cdot 10^{11}$~cm$^{-2}$, $m^*=0.2 m_0$ and spin
splitting constant
$\beta=5$~meV$\cdot$nm~\cite{Winkler03,Schneider2004}, we obtain
an average spin polarization of $3.2 \cdot 10^{-4}$ and $0.8 \cdot
10^{-4}$ for the experimentally relevant current densities. Since
the values obtained from an analysis of our data under the
assumption of dominating dichroic absorption is by a factor of
more than 1000 higher than expected  we assume that Faraday
rotation and not dichroic absorption dominates the change of
polarization of the transmitted light. Also the fact that the spin
orientation induced signal increases linearly with current (see
Fig.~\ref{fig03}d ) and not quadratically, as expected from
dichroic mechanism (see  Eqs.~(\ref{dichro,Srel})), points to the
Faraday rotation as the dominating mechanism proofing current
induced spin orientation.

So far we assumed that  the subband spin splitting occurs for spin
eigenstates pointing normal to the QW. However, if the hole
subbands are also split   due to a spin-orbit coupling $\propto
\sigma_x k_y$ in the Hamiltonian an additional mechanism of spin
orientation, the precessional
mechanism~\cite{aronov89,Aronov91p973}, needs to be taken into
account. The difference in the spin relaxation rates for spin-up
and spin-down subbands  are now determined by the D'yakonov-Perel
spin relaxation process.
 In this case the relaxation rate depends
on the average $\bm k$-vector~\cite{Dyakonov86p110},  equal to
${\bar k}_{3/2}= - k_0 + \langle k \rangle$ for the spin-up
 and ${\bar k}_{-3/2}= k_0 + \langle k \rangle$ for the
spin-down subband. Hence also for the D'yakonov-Perel spin
relaxation mechanism a current through the hole gas causes spin
orientation. If this type of spin-orbit interaction is present,
the magnitude of spin orientation is also given by Eq.~\ref{Srel},
only the constant Q is different but also of order
1~\cite{Aronov91p973}.

Finally,  we discuss our results in the light of related
experiments. Based on theoretical predictions made by Ivchenko and
Pikus~\cite{Ivchenko78p640},  Vorob'ev et al. observed a current
induced spin polarization in bulk tellurium~\cite{Vorobjev79p441}.
This is a consequence of the unique band structure of tellurium
with hybridized spin-up and spin-down bands and is, other than in
our experiment, not related to spin relaxation. More recently for
spin injection from a ferromagnetic film into a two-dimensional
electron gas Hammar et al. used the above concept of a spin
orientation by current in a 2DEG~\cite{Hammar99p203} (see
also~\cite{Johnson98p9635,Silsbee01p155305}) to interpret their
results. Though a  larger degree of spin polarization was
extracted  the experiment's interpretation is complicated by other
effects~\cite{Monzon00p5022,Wees00p5023}. We would also like to
note that Kalevich and Korenev~\cite{kalevich90} reported an
influence of an electric current on the spin polarization achieved
by optical orientation. The current does not align spins, but the
effective magnetic field due to the current causes a spin
depolarization like the Hanle effect in an external magnetic
field. While preparing the manuscript we became aware of
experimental results obtained on strained InGaAs bulk
material~\cite{Kato2004}. Analyzing Faraday rotation the authors
of this preprint also report on the build up of a spin
polarization under current bias, however, in three dimensional
system.

We thank  E.L.~Ivchenko, V.A.~Shalygin and A.Ya.~Shul'man for
helpful discussions. We acknowledge financial support from the
DFG, RFBR, INTAS, and ``Dynasty'' Foundation.

\end{document}